\documentclass[twocolumn,twoside,preprintnumbers,amsmath,amssymb,showkeys]{revtex4}
\usepackage{epsfig}
\usepackage{graphicx}
\usepackage{dcolumn}
\usepackage{bm}
\usepackage{fancyhdr}
\usepackage{pslatex}

\begin{document}

\title{{  Strong interactions and gauge/string duality   }}

\author{Henrique Boschi-Filho and Nelson R. F. Braga}

\affiliation{Instituto de F\'{\i}sica, 
Universidade Federal do Rio de Janeiro, Caixa Postal 68528, 
RJ 21941-972 -- Brazil}


\begin{abstract}
We discuss some recent phenomenological models for strong interactions based on the idea of
gauge/string duality.
A very good estimate for hadronic masses can be found by placing an infrared cut off in AdS space.
Considering static strings in this geometry one can also reproduce the phenomenological Cornell potential 
for a quark anti-quark pair at zero temperature. Placing  static strings in an AdS Schwarzschild space with an infrared cut off one finds a transition from a confining to a deconfining phase at some critical 
horizon radius (associated with temperature).  


\end{abstract}

\maketitle

\section{Introduction}

In this review we are going to discuss some recent phenomenological results 
concerning the relation between string theory and QCD
inspired in this idea of introducing an infrared cut off in anti-de Sitter (AdS) or in Schwarzschild-AdS spaces.
QCD has been tested and confirmed with success in high energy 
experiments but it is non perturbative at low energies.
Lattice calculations give us very important results in this regime.
However it seems that we are still far from a complete description of the 
complexity of strong interactions. In particular, important aspects 
like confinement and mass generation still lack a satisfactory description.
Presently there are many indications that string theory can be useful
in the description of strong interactions in the non perturbative regime  
of QCD.
 
An early connection between SU(N) gauge theories (for large N) and string theory
was realized long ago by 't Hooft\cite{'tHooft:1973jz}.
A few years ago a very important result was obtained by Maldacena\cite{Malda}.
He established a correspondence between string theory in $AdS_5 \times S^5 \,$ 
space-time and ${\cal N } = 4$ superconformal Yang Mills SU(N) theory for large N 
at its four dimensional boundary,  known as AdS/CFT 
correspondence\cite{Malda,GKP,Wi}. 

In the AdS/CFT correspondence there is an exact duality between a four dimensional gauge theory and string theory in a ten dimensional space.
However, in this formulation, the gauge theory has no energy scale as it is conformal.  
Although it involves a conformal gauge theory, the AdS/CFT correspondence has been a very important source of inspiration for searching QCD results from string theory.
The first idea of breaking conformal invariance in the AdS/CFT context,
proposed by Witten, is to consider an AdS Schwarzschild black hole as dual to a non-supersymmetric Yang Mills theory\cite{Wi2}. This approach was used to
calculate glueball masses in \cite{MASSG,MASSG2,MASSG3,MASSG4,MASSG5,MASSG6,MASSG7}.

\section{QCD scattering and string theory}

A very important result relating string theory to the behavior of scattering amplitudes
for high energy processes at fixed angles was found by Polchinski and Strassler\cite{PS}. 
In this regime, that corresponds, in terms of Mandelstam variables,  to $ s \rightarrow \infty \,\,\, $ with $\,\,s/t\,$ fixed, the Veneziano amplitude coming from string theory in flat space shows a soft scattering behavior

\begin{equation}
\,\,\, A_{_{Ven.}} \sim exp\,\{\, -\alpha^\prime s f (\theta)\, \} \,,
\end{equation}
 
\noindent where $\theta $ is the scattering angle and $\alpha^\prime $ is related to the string tension.
In contrast, it was known for a long time that hadronic scattering amplitudes for processes in this
regime show a hard scattering behavior, as reproduced by QCD\cite{QCD1,BRO}. 
That means, the amplitudes fall of with a negative power of $s$.  
Polchinski and Strassler found a solution to this apparent obstacle in 
the description of strong interactions by string theory considering 
the duality between gauge theory glueballs and string theory dilatons in an AdS 
space with an infrared cut off. This way they found the 
QCD hard scattering behaviour for high energy amplitudes at fixed angles.
 
The hard scattering behavior was also obtained afterwards in \cite{BB3} from a  mapping between quantum states 
in AdS space and its boundary found in \cite{BB2}. 
We considered an AdS slice as approximately dual to a confining gauge 
theory.  The slice corresponds to the metric

\begin{equation}
\label{AdSPoincare}
ds^2=\frac {R^2 }{( z )^2}\Big( dz^2 \,+(d\vec x)^2\,
- dt^2 \,\Big)\,,
\end{equation}

\noindent with $\, 0\le z \le  z_{max} \,\, \sim 1/\mu $  where  $\mu$
is an energy scale chosen as the mass of the lightest glueball.
We used a mapping between Fock spaces of a scalar field in AdS space and operators 
on the four dimensional boundary, defined in \cite{BB2}. 
Considering a scattering of two particles into $m$ particles 
one finds a relation between bulk and boundary scattering amplitudes\cite{BB3}

\begin{equation}
S_{Bulk} \, \sim \,  
 S_{Bound.} \,\,\Big( {\sqrt{\alpha^\prime} \over \mu }\Big)^{m+2} \,\, K^{(m+2)(1+d)}
\end{equation}

\noindent where $d $ is the scaling dimension of the boundary operators and $K$ is the boundary momentum scale. This leads to the result for the amplitude

\begin{equation}
A_{Boundary} \,\sim \,s^{(4  - \Delta)/2 } \,,
\end{equation}

\noindent where $\Delta$ is the total scaling dimension 
of scattered particles. This reproduces the hard scattering behavior.

For some other results concerning QCD scattering properties from string theory see also 
\cite{GI,BT,AN,PS2,Brodsky:2003px,AN2,AN3}. 
 
\section{Scalar glueball masses }

Using the phenomenological approach of introducing an energy scale 
by considering an AdS slice we found estimates for scalar glueball mass ratios\cite{BB4,BB5}.
In the AdS$_5$ bulk we took dilaton fields satisfying Dirichlet boundary conditions at $ z = z_{max}$ 
\begin{eqnarray}
\label{QF}
\Phi(z,\vec x,t) &=& \sum_{p=1}^\infty \,
\int { d^3 k \over (2\pi)^{3}}\,
{z^{2} \,J_2 (u_p z ) \over z_{max}\,\, w_p(\vec k ) 
\,J_{3} (u_p z_{max} ) }\nonumber\\
&\times& \lbrace { {\bf a}_p(\vec k )\ }
 e^{-iw_p(\vec k ) t +i\vec k \cdot \vec x}\,
\,+\,\,h.c.\rbrace\,,\nonumber
\end{eqnarray}

\noindent where $\,w_p(\vec k ) \,=\,
\sqrt{ u_p^2\,+\,{\vec k}^2}\,$, 

\begin{equation}
 u_p \,=\,\frac{ \chi_{_{2\,,\,p}}}{z_{max}} \,,
\end{equation}

\noindent is the momentum associated with the $z$ direction
and $\chi_{_{2\,,\,p}} $ are the zeroes of the Bessel functions:  
$\,J_2 (\chi_{_{2\,,\,p}} )=0\,.$
 
 On the boundary ($ z = 0)$ we considered scalar glueball states  $J^{PC}\,=\,0^{++}$ 
and their excitations $0^{++\ast},\,\,0^{++\ast\ast}$  with masses 
$\mu_p\,,\,p=1,2,...$. 
Assuming an approximate gauge/string duality the 
glueball masses are taken proportional to the dilaton discrete modes: 
$$
{u_p \over \mu_p }\,=\,{\rm const.}\,
$$

\noindent So, the ratios of glueball masses are related  
to zeroes of the Bessel functions
$$
{ \mu_p\over \mu_1 }\,=\,{\chi_{2\,,\,p}\over \chi_{2\,,\,1}}\,\,.
$$

\bigskip

\noindent Note that these ratios are independent of the size of the slice   $\,\,\,\,\,z_{max}\,$. 
Our estimates are in good agreement with the available lattice\cite{LAT1,LAT2} and AdS-Schwarzschild 
\cite{MASSG} results.  For a detailed comparision see refs. \cite{BB4,BB5}.
 
For some other results concerning glueball masses using gauge/string duality
see for instance \cite{Caceres:2000qe,ACEP,Amador:2004pz,Schvellinger:2004am,Caceres:2005yx}.

\section{ Higher spin states and Regge trajectories}

Recently, very interesting results for the hadronic
spectrum were obtained by de Teramond and Brodsky\cite{deTeramond:2005su} considering scalar, vector and fermionic fields in a sliced $ AdS_5 \times S^5 $ space.
It was proposed that massive  bulk states corresponding to fluctuations about 
the $AdS_5$ metric are dual to QCD states with  
angular momenta (spin) on the four dimensional boundary.
This way the spectrum of light baryons and mesons has been reproduced
from a holographic dual to QCD inspired by the AdS/CFT 
correspondence.

We used a similar approach to estimate masses of glueball states 
with different spins\cite{Boschi-Filho:2005yh}. 
The motivation was to compare the glueball Regge trajectories with
the pomeron trajectories. For soft pomerons \cite{Landshoff:2001pp} experimental
results show that the spin $J$ of the pomeron is
\begin{equation}
\label{11}
J \,\approx \,  1.08 \,+\, 0.25\, M^2\,\,\,\,\,\,\,\,\,\,,
\end{equation}

\noindent where $M$ is the mass in GeV. It is conjectured that the soft pomerons may be related to glueballs.
Recent lattice results are consistent with this interpretation\cite{Meyer:2004jc}.
 
We assume that massive scalars in the AdS slice with mass $\mu$ 
are dual to boundary gauge theory states with spin $J$ related by: 
\begin{equation}
( \mu R )^2 \,=\, J ( J + 4 ) \,\,.
\end{equation}

We consider both Dirichlet and Neumann boundary conditions and the results for the
four dimensional glueball masses with even spin are shown in tables I and II respectively.

\begin{table}
\begin{tabular}{ | c | c | c | c |} 
\hline 
Dirichlet $\,\,$ 
 & $\,\,$ lightest $\,\,$   &
$1^{st}$ excited $\,\,$ & $2^{nd}$ excited \\
glueballs $\,\,$     &   state & state & state \\
 \hline
 $0^{++}$ & 1.63  &  2.67 & 3.69 \\ 
 $2^{++}$ & 2.41    & 3.51  & 4.56  \\
 $4^{++}$ & 3.15  & 4.31  & 5.40  \\
 $6^{++}$ & 3.88  & 5.85  & 6.21 \\
 $8^{++}$ & 4.59 & 5.85  & 7.00 \\
 $10^{++}$ & 5.30 & 6.60   & 7.77 \\
\hline
\end{tabular}
\caption{     Higher spin glueball masses in GeV with Dirichlet boundary condition. 
The value 1.63 is an input from lattice.}
\end{table}

\begin{table}
\begin{tabular}{ | c | c | c | c |} 
\hline 
Neumann $\,\,$ 
 & $\,\,$ lightest $\,\,$   &
$1^{st}$ excited $\,\,$ & $2^{nd}$ excited \\
glueballs $\,\,$     &   state & state & state \\
 \hline
 $0^{++}$ & 1.63  & 2.98  & 4.33  \\ 
 $2^{++}$ & 2.54    & 4.06 & 5.47  \\
 $4^{++}$ & 3.45 & 5.09  & 6.56 \\
 $6^{++}$ & 4.34  & 6.09 & 7.62 \\
 $8^{++}$ & 5.23 & 7.08 & 8.66 \\
 $10^{++}$ & 6.12 & 8.05  & 9.68 \\
\hline
\end{tabular}
\caption{  Higher spin glueball masses in GeV with Neumann boundary condition. 
The value 1.63 is an input from lattice.  }
\end{table} 
   
We found non linear relations between spin and mass squared.
We considered linear approximations representing Regge trajectories
\begin{equation}
 J \,=\, \alpha_0 \,+\,\alpha^{\prime} \, M^2\,.
\end{equation}

\begin{figure}[htbp]
\begin{center}
\includegraphics[width=8cm]{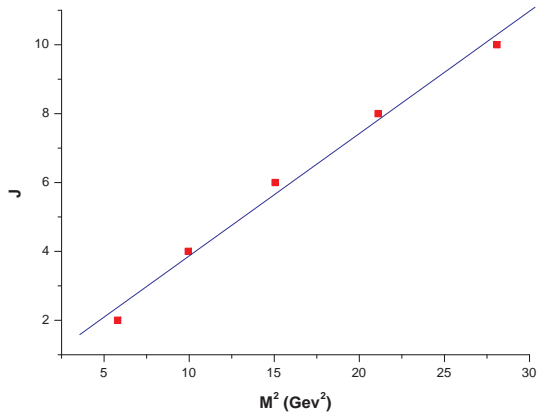}
\caption{ Spin versus mass squared for the lightest glueball states with 
 Dirichlet boundary conditions from table I. 
The line corresponds to the linear fit. }
\end{center}
\end{figure}

\noindent For Dirichlet boundary conditions, taking the states 
$J^{++}\,$ with $J = 2,4,...,10\,$ 
we found a linear fit with 
\begin{equation}
\alpha^{\prime}\,=\,( \, 0.36 \pm  \,0.02\,)\,GeV^{-2}\, \quad ; \quad
 \alpha_0 \,=\,0.32 \pm 0.36 \,,
\end{equation}

\noindent as shown in Figure 1.

\noindent For Neumann boundary conditions for the states  
$J^{++}\,$ with $J = 2,4,...,10\,$ we found 
\begin{equation}
\alpha^{\prime}\,=\,(\,0.26 \pm 0.02 \,)GeV^{-2} 
\quad ; \quad \alpha_0 \,=\,0.80 \pm 0.40\,,
\end{equation}

\noindent as shown in Figure 2.

\begin{figure}[htbp]
\begin{center}
\includegraphics[width=8cm]{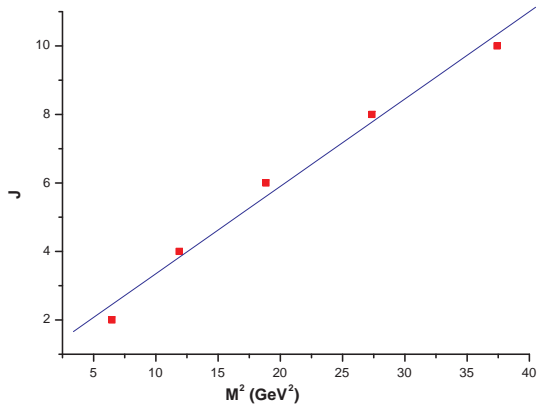}
\caption{  Spin versus mass squared for the lightest glueball states with 
Neumann boundary conditions from table II. The line corresponds to the linear fit.}
\end{center}
\end{figure}

So, comparing these results with eq. (\ref{11}) we see that Neumann boundary conditions give a glueball trajectory consistent with that of pomerons. These kind of boundary conditions appear in the Randall Sundrum 
model\cite{Randall:1999ee} as a consequence of the orbifold condition.
    
\section{Wilson loops and quark anti-quark potential}

Wilson loops are an important tool to discuss confinement in gauge theories since they 
show the behavior of the energy associated with a given field configuration.
In the AdS/CFT correspondence Wilson loops for a heavy quark anti-quark pair in the
gauge theory can be calculated from a static string in the 
AdS space\cite{RY,MaldaPRL}. The corresponding energy is a non confining Coulomb
potential as expected for a conformal theory. For an excellent review and extension to other metrics
see \cite{Kinar:1998vq}.  
We calculated Wilson loops for a quark anti-quark pair in D3-brane space finding different behaviors, with respect to confinement, depending on the quark position\cite{Boschi-Filho:2004ci}.

The static potential energy of a heavy quark anti-quark pair   
is described by the phenomenological Cornell potential 

\begin{equation}
\label{Cornell}
E_{Cornell}(L) \,=\, -\frac{4}{3} \frac{a}{L} \,+\, \sigma L\,+\, constant\,\,,
\end{equation}
\noindent where $a = 0.39\,$ and $\sigma = 0.182 GeV^2\,$. 

The metric (\ref{AdSPoincare}) of the AdS space can be rewritten as 

\begin{equation}
\label{metric}
ds^2 \,=\, \Big( {r^2\over R^2} \Big) ( -dt^2 + d{\vec x}^2 ) +  
\Big( {R^2\over r^2} \Big) dr^2 \,,
\end{equation}

\noindent where $\, r = R^2/z $. We have calculated the energy of a static string in an AdS slice defined by 
 $ r_2 \le r \le r_1 \,$ \cite{Boschi-Filho:2005mw}. 
The quark anti-quark pair (string endpoints) is located 
at $r = r_1 $, separated by a four dimensional ($x$ coordinates) distance $L$  and there is 
an infrared cut off in the space at $ r= r_2$. From now on we choose $r_2 \,=\, R$. 
Note that there are two kinds of geodesics, as shown in figure 3, depending on the value of $L$. 
For small quark separation $ L \le L_{crit}\,$ the geodesics are curve (like curve {\bf a} ) with one 
minimum value of the coordinate $r = r_0 $ which is related to $L$ by

\begin{equation}
\label{Lr1}
L (r_0 ) \,=\,\frac{ 2 R^2 }{r_0}\,I_1 (r_1/r_0 )
\end{equation}

\noindent where $I_1 (\xi )$ is the elliptic integral
\begin{equation}
I_1 (\xi )\,=\,\int_1^{\,\xi}\,
\frac{ d \rho }{\rho^2 \,\sqrt{ \rho^4 -1}}\,.
\end{equation}

\noindent The critical value corresponds to $ L_{crit} = L ( r_0 = R ) $ as in curve {\bf b} of figure 3.  

The energy for $L \le L_{crit}\,$ can be calculated as

\begin{equation}
\label{E-}
E^{\,(-)} \,=\, \frac{ 2 R^2 }{\pi \alpha^\prime} \,\frac{I_1 (r_1/r_0 )}{L}
\Big[ \, I_2 (r_1/r_0)\,-\,1 \Big] \,.
\end{equation}

\noindent where $1/2\pi \alpha^\prime $ is the string tension and we have subtracted the 
constant $ r_1 /\pi \alpha^\prime\,$  in such a way that the energy is finite 
even in the limit $r_1 \to \infty $. The integral $I_2 $ is  

 \begin{equation}
I_2 (\xi) \,=\, \int_1^{\xi}\,
\Big[\,\frac{ \rho^2 }{\sqrt{ \rho^4 -1}} \,-\,1 \,\Big] d\rho \,.
\end{equation}

For $ L > L_{crit} $, the geodesics reach the infrared brane as shown in curve {\bf c} of figure 3. 
The energy can be calculated again subtracting the constant $ r_1 /\pi \alpha^\prime\,,$ associated with the quark mass. 
We obtain

\begin{eqnarray}
\label{Er1}
E^{\,(+)} &=&  \frac{ 2 R^2 }{ \,\pi \alpha^\prime} \frac {I_1 ( r_1/r_0)}{L}
\Big[ I_2( r_1/r_0) - I_2 (R/r_0)\Big]
\,-\,\frac{ R }{\pi \alpha^\prime} \nonumber\\
&+& 
 \frac{ L  }{2 \pi \alpha^\prime\,} 
\frac {I_1 ( R/r_0)}{ I_1 (r_1/r_0)} \,,
\end{eqnarray}

\begin{figure}
\centering

\
\setlength{\unitlength}{0.07in}
\vskip 5.5cm
{\begin{picture}(0,0)(18,0)
\rm\thicklines\bf
\put(25,-8){\vector(0,0){38}}
\put(26,-0.5){$- L_{crit}/2$}
\put(26,19.5){$+L_{crit}/2$}
\put(26,30){$x$}
\put(24.5,-0.5){$\bullet$}
\put(24.5,19.5){$\bullet$}
\put(24.5,14.5){$\bullet$}
\put(24.5,4.5){$\bullet$}
\put(24.5,24.5){$\bullet$}
\put(24.5,-5.5){$\bullet$}
\put(17,18){b}
\put(19,14){a}
\put(19,24){c}
\put(3,10){\vector(1,0){30}}
\put(33,8){$r $}
\put(22.7,-10 ){$r = r_1 $}
\put(10,-6){$r = r_2$}
\bezier{600}(25,15)(10,10)(25,5)
\bezier{600}(25,20)(0,10)(25,0)
\bezier{600}(25,25)(17,22)(12.6,18)
\bezier{600}(25,-5)(17,-2)(12.6,2)
\bezier{600}(12.5,18)(12.5,10)(12.5,2)
\bezier{600}(12.55,18)(12.55,10)(12.55,2)
\multiput(12.55,-4.3)(0,1){30}{\line(0,1){0.5}}
\end{picture}}
\vskip 2.5cm
{\caption{ Schematic representation 
of geodesics in the AdS slice. }}
\end{figure}
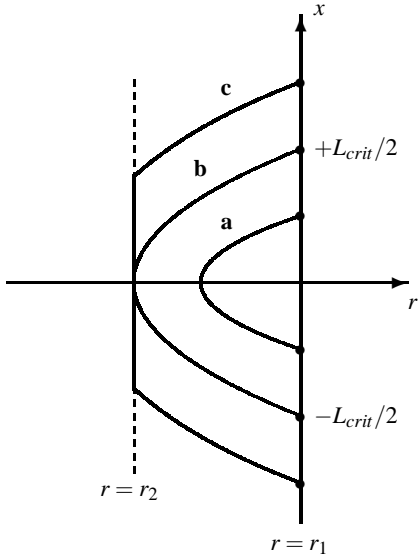

Taking the limit $ r_1 \to \infty \,$, we 
find a potential which is approximately Coulombian for small $ L $ and 
has a leading linear confining behavior for  large distances

\begin{equation}
 E^{\,(+)} \,\sim \, \frac{1}{2\pi\alpha'} \, L \,.
\end{equation}

\noindent The identification of this potential energy with the Cornell potential leads to 
  
\begin{equation}
a \,=\, 3 C_1^2 R^2/2\pi \alpha'\,\,\,\,\,;\,\,\,\,\,\,\,
\sigma \,=\, \frac{1}{2\pi\alpha'}  
\end{equation} 
  
\noindent with $C_1\,= \,\sqrt{2}\pi^{3/2}/[\Gamma(1/4)]^2\,\,.$ So that we find an
effective AdS radius $ R = 1.4 \,{ \rm GeV}^{-1}\,$. 
Then the energy takes the form
 
\begin{eqnarray} 
E^{\,(-)} &=&
\,-\, \frac{ 4 a }{ 3 \,L } \,
\,,  \qquad L \le L_{crit} \\ 
 E^{\,(+)} &=& - \frac{ 4 a }{ 3 L}
\,+\,\frac{ 4 a }{ 3 C_1 \,L} 
\Big[ 1 - I_2 (R/r_0)\Big] \nonumber\\
&-& \sqrt{\frac{\, 4 \,a \,\sigma }{ 3\, C_1^2}} \,+\,
 \sigma  L \,\,\frac {I_1 ( R/r_0)}{ C_1 }\,, \qquad L \ge L_{crit}
\end{eqnarray}

The shape of this potential energy is very similar to the Cornell potential as shown in figure 4.

\begin{figure}
\centering
\includegraphics[width=8cm]{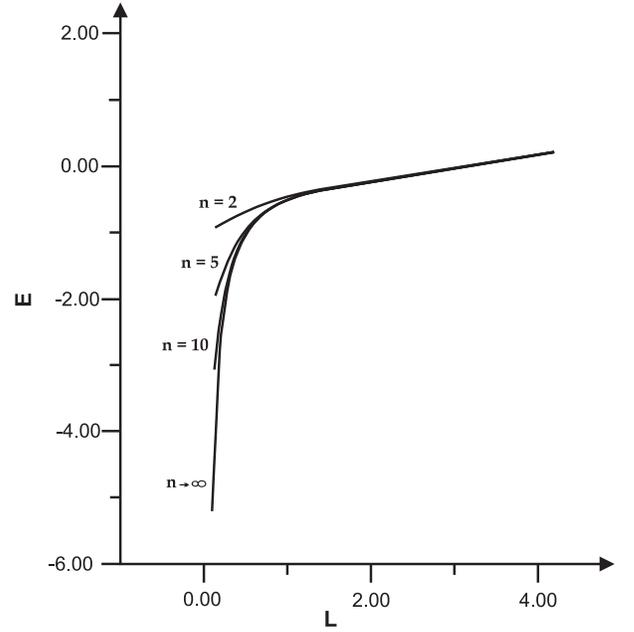}
{\caption{ Energy in GeV as a function of string end-points 
separation $L$ in GeV$^{-1}$, for AdS slices with
 $r_1 = nR $ and $r_2 = R $ . For $ n \to \infty$  
the energy behaves as the Cornell potential eq. (\ref{Cornell}).}}
\end{figure}
 
\section{Quark anti-quark potential at finite temperature}  

A gauge/string duality involving a gauge theory at finite temperature 
was proposed by Witten in \cite{Wi2} inspired by the work of
Hawking and Page\cite{Hawking:1982dh}.
In this approach, for high temperatures, the AdS space accommodates a  Schwarzschild black hole and the
horizon radius is proportional to the temperature. For low temperatures the dual space would be an AdS space with compactified time dimension, known as thermal AdS. 
The gauge string duality using AdS Schwarzschild space has recently been applied to obtain the viscosity of a quark gluon plasma\cite{Policastro:2001yc,Kovtun:2004de}. 

In ref. \cite{Boschi-Filho:2006pe} we considered an AdS Schwarzschild black hole metric as a phenomenological model for a space dual to a theory with both mass scale and finite temperature. 
The corresponding metric is

\begin{equation}
\label{metric}
ds^2 \,=\, \Big( {r^2\over R^2} \Big) ( -\, f(r) \,dt^2 \,+ \, d{\vec x}^2 \,) +  
\Big( {R^2\over r^2} \Big) \,\frac{1}{f(r) } \, dr^2 \,+\, R^2 d^2 \Omega_5\,,
\end{equation}

\noindent where $r_2 \le r < \infty \,\,$, $\,\,\,  f(r)\,=\, 1\,-\, r_T^4 / r^4 \,\ $ 
and the horizon radius $r_T\,$ 
is related to the Hawking temperature by $r_T\,=\, \pi\, R^2 \,T\,$. At zero temperature this space becomes an  Anti-de Sitter (AdS) slice. 
The problem of static strings in a space with metric (\ref{metric}), without any  cut off,
was discussed in detail in\cite{Rey:1998bq,Brandhuber:1998bs}. 

Calculating the energy of static strings in this space we found a deconfinement phase transition at a 
critical temperature. This transition shows up because, depending on the horizon radius (temperature) relative to the cut off position $r_2$, we have different behaviors for the energy. 
If $r_T \ge r_2$ the  string will not be affected by the presence of the brane since it does not cross
the horizon. So the energy will be that described in ref. \cite{Rey:1998bq,Brandhuber:1998bs}
and there will be no confinement. If $r_T <  r_2$ the energy for large endpoint separation $L$ 
will grow linearly with $L$ with a temperature dependent coefficient and the quarks are confined. 
The critical temperature $\, T_C\,$ corresponds to $r_T \,=\, r_2$. 

As in the zero temperature case we choose $r_2 = R$ from now on. 
We show in figure 5 the energies obtained in \cite{Boschi-Filho:2006pe}
for temperatures $ T =0,\, T = 0.8 T_C \,,\, T = T_C\,$ and $ \,T = 2 T_C\,$. 
This figure illustrates the fact that in our model the energy of static strings associated with the quark anti-quark potential present a confining behavior for 
temperatures below  $ T_C \,=\, 1 /\, \pi\, R\,$. In this case there is a linear term in the energy, for large quark distances $L\,\,$ given by  $\,E \sim  \sigma (T)\,L\,$ with 

\begin{equation}
\label{tensionT}
\sigma (T) \,=\,\frac{1}{2 \pi \alpha^\prime} \sqrt{ 1 -  (\pi R T )^4 }\,\,\hskip1cm ( T < T_C ).
\end{equation}
 
\noindent At zero temperature this coefficient is identified with the string tension of the Cornell potential 
$ 1/ ( 2 \pi \alpha^\prime ) \,=\, 0.182$ Gev$^2$. 
For temperatures $ T \ge  T_C $ there is no confinement since the energy is finite 
when $L \to \infty$.   
Choosing the brane position to have the value $ R = 1.4 \,{ \rm GeV}^{-1}\,$ 
as in the zero temperature case\cite{Boschi-Filho:2005mw} we find a critical temperature $ T_C \,\sim \,{\rm  230 MeV}\,$.

\begin{figure}
\centering
\includegraphics[width=8cm]{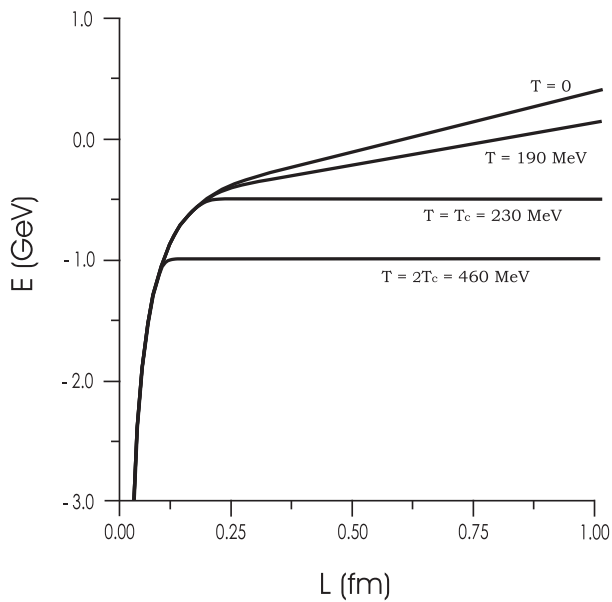}
{\caption{ Energy as a function of string end-points separation for different temperatures. } }
\end{figure}

Our results agree qualitatively with lattice calculations for QCD at finite 
temperature \cite{Kaczmarek:1999mm,Kaczmarek:2002mc,Kaczmarek:2004gv,Petreczky:2005bd}.  
However, it is important to mention that the results from fluctuations of strings in flat space at  finite temperature\cite{Pisarski:1982cn} and from lattice calculations\cite{deForcrand:1984cz} imply corrections to the string tension of order $\,-\, T^2\,$ at low temperatures, while our model predicts corrections of order $\,- \,T^4 \,\, $, as can be seen from  eq. (\ref{tensionT}). 
If we had considered the thermal AdS metric for low temperatures, instead of the AdS Schwarzschild black hole metric, we would get no thermal corrections to the string tension. It would be interesting to find
a holographic phenomenological model that gives the expected low temperature corrections.  
It is worth mentioning the recent articles  \cite{Ghoroku:2005kg} and \cite{Andreev:2006eh} 
that also discuss thermal effects in the gauge/string duality context.
There is also a very recent result by Herzog\cite{Herzog:2006ra} which indicates that the dual space 
for temperatures bellow $T_C$ should be the thermal AdS.  

For other interesting results concerning gauge/string duality and QCD see for 
instance  \cite{Janik:2001sc, PandoZayas:2003yb,Andreev:2004sy,Bigazzi:2004ze,Erlich:2005qh,
DaRold:2005zs,Evans:2005ip,Hambye:2005up,Casero:2006pt,Brodsky:2005kc,Brodsky:2006uq}.

\acknowledgements{ We would thank Silvio Sorella and the organizers of ``Infrared QCD in Rio",  where this talk was presented, for the very warm atmosphere during the workshop. We thank also Michael Teper and Andrei Starinets for very interesting discussions about the finite temperature case. The authors are partially supported by CNPq and Faperj.}

\end{document}